\documentclass[%
 reprint,
 amsmath,amssymb,
 aps,
]{revtex4-2}

\usepackage{xcolor}

\usepackage{graphicx}% Include figure files
\usepackage{dcolumn}% Align table columns on decimal point
\usepackage{bm}% bold math
\usepackage{upgreek} % for non-italic greek letters

\begin{document}

\preprint{APS/123-QED}

\title{Lifetime measurement of the $5s5p~^1P_1$ state in strontium}% Force line breaks with \\

\author{Ivana Puljić}
\author{Ana Cipriš}\altaffiliation{Present address: Instituto de Física de São Carlos, Universidade de São Paulo, São Carlos, SP 13566-970, Brazil}
\author{Damir Aumiler}
\author{Ticijana Ban}
\author{Neven Šantić}
\email{nsantic@ifs.hr}
\affiliation{Institute of Physics, Centre for Advanced Laser Techniques, Bijenička cesta 46, 10000 Zagreb, Croatia }


\date{\today}% It is always \today, today,

\begin{abstract}
We present a direct lifetime measurement of the $5s5p~^1P_1$ state of strontium using time-correlated single-photon counting of laser induced fluorescence in a hot atomic beam. 
To achieve fast switch-off times and a high signal-to-noise ratio, we excite the strontium atoms with a femtosecond pulsed laser at $\approx$461 nm and collect the fluorescence onto a hybrid single-photon detector. Analysis of the measured exponential decay gives a lifetime of the $^1P_1$ state of $\tau = (5.216 \pm 0.006_{stat} \pm 0.013_{sys})$~ns, where all the systematic effects have been thoroughly considered. 
% \begin{description}
% Secondary publications and information retrieval purposes.
% You may use the \texttt{description} environment to structure your abstract;
% use the optional argument of the \verb+\item+ command to give the category of each item. 
% \end{description}
\end{abstract}

\maketitle

%\tableofcontents

\section{\label{sec:Introduction}Introduction\protect\\ }%The line
%break was forced \lowercase{via} \textbackslash\textbackslash}

In the last two decades, strontium optical lattice clocks have continuously advanced the frontier in lowering accuracy uncertainty, with recent realizations reaching the 10$^{-19}$ range \cite{Aeppli2024}.
These levels of accuracy not only put Sr as one of the leading candidate elements for the redefinition of the second \cite{Dimarcq2024}, but also promise to advance geodesy \cite{Mehlstäubler2018, Grotti2018, Grotti2024, Takamoto2020, Huang2020, Bothwell2022, Zheng2023}, impose constraints on dark matter models \cite{Derevianko2014, Arvanitaki2015, Beloy2021, Sherrill2023, Filzinger2023} and on the variation of fundamental constants \cite{Safronova2018, Barontini2022}.

The largest contribution to the uncertainty of state-of-the art Sr lattice clocks comes from the black body radiation (BBR) shift. 
This shift arises from considerable differential polarizability of the clock states, $5s{^2}~^1S_0$ and $5s5p~^3P_0$, at wavelengths corresponding to the thermal radiation at room temperature at which these clocks typically operate. 
The static part of the polarizability, which scales as $T^4$, has been measured to  such a high precision \cite{Middelmann2012} that it contributes negligibly to the overall uncertainty. 
However, the dynamic part, which scales with higher powers of $T$, needs to be calculated from the available spectroscopic data and is the primary contributor to the BBR shift uncertainty \cite{Safronova2013, Nicholson2015,Hobson2020,Lisdat2021,Aeppli2024}. 
This dynamic polarizability depends on the Einstein coefficients $A$ between higher-lying states and the clock states.
For the $5s{^2}~^1S_0$ state the the dominant contribution - over 90\% of the total polarizability, comes from the $5s5p~^1P_1$ state. 
Therefore, a precise and accurate determination of $A(5s5p~^1P_1 \rightarrow 5s^2~^1S_0)$ is crucial for lowering the uncertainty of Sr lattice clocks.

The commonly used value was measured through photoassociation spectroscopy \cite{Katori2006}. 
Alternatively, it can be calculated from the combination of measurements of the tune-out wavelength \cite{Blatt2020} and the lifetime of the $^3P_1$ state \cite{Ye2015, Aeppli2024}. 
However, these two approaches show a $7\sigma$ disparity, an inconsistency also noted in \cite{Lisdat2021}.

In this paper, we report on the first measurement of the lifetime of the $5s5p~^1P_1$ state in atomic Sr using femtosecond laser excitation and time-correlated single photon counting (TCSPC). 
As a direct measurement of the lifetime, our result has the potential to resolve the previously mentioned $7\sigma$ disparity and contribute to a more precise calculation of the dynamic polarizability correction. 
This, in turn, will enable a more accurate calculation of the BBR shift, ultimately improving the precision of Sr lattice atomic clocks.

In our experiment, we generate an atomic beam in an under-vacuum spectroscopy cell using a dispenser, excite it with a femtosecond (fs) laser resonant with the $5s^2~^1S_0 \rightarrow 5s5p~^1P_1$ transition at $\approx~461$~nm, and collect the fluorescence onto a hybrid single-photon detector.

By using the TCSPC method, we were able to measure individual photons emitted by atoms following the excitation with a fs laser and record their arrival times with picosecond resolution \cite{Becker2023}. 
Over many cycles of excitation and spontaneous emission, an exponential histogram of photon counts in time is recorded, with the time constant being the lifetime of the excited state. 

Using a fs laser emitting at $\approx461$~nm, we achieve the necessary fast switch-off times shorter than the $5s5p~^1P_1$ lifetime of $\approx$5~ns. 
This approach allowed us to bypass the need for electro-optic modulators, which are commonly used for fast switch-off of laser excitation at red and near-infrared wavelengths \cite{Araujo2016}, but are unavailable for blue wavelengths, thus making the measurements of short lifetimes in the blue spectrum technically challenging.

Our paper is organized as follows. Details of the experiment are given in Section \ref{sec:Exp setup} and our approach to the data analysis in Section \ref{sec: Analysis}. In Sections \ref{sec: Lifetime} and \ref{sec: Systematic effects} we present our data and discuss all the systematic effects contributing to the error budget of our measurement. Finally, we conclude in Section \ref{sec: Conclusion}.

%%%%%%%%%%%%%%%%%%%%%%%%%%%%%%%%%%%%%%%%%%%%%%%%%%%%%%%%%%%%%%%%%%%%%%%%%%%%%%%%%%%%%%%%%%%%%%%%%%%%%%%%%%%%%%%%%%%%%%%%%%%%%%%%
%%%%%%%%%%%%%%%%%%%%%%%%%%%%%%%%%%%%%%%%%%%%%%%%%%%%%%%%%%%%%%%%%%%%%%%%%%%%%%%%%%%%%%%%%%%%%%%%%%%%%%%%%%%%%%%%%%%%%%%%%%%%%%%%

\section{\label{sec:Exp setup}Experimental methods\protect\\} \label{sec: exp}

\begin{figure}[h]
\includegraphics[width = \columnwidth]{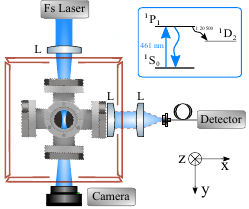}% Here is how to import EPS art
\caption{\label{fig:scheme} A simplified scheme of the experimental setup and relevant energy levels of strontium. The atomic beam of strontium atoms is propagating in -z direction and is produced using a dispenser in the center of the vacuum cell.}
\end{figure}

A simplified scheme of the experimental setup and energy levels of strontium relevant to the measurement of the $^1P_1$ lifetime is shown in Fig.~\ref{fig:scheme}. 
Even though there is a probability of atom relaxation from the $^1P_1$ state to the $^1D_2$ state, this occurs only once per 20500 photons \cite{Cooper2018} and is therefore negligible.

An atomic beam of hot strontium atoms is generated using a dispenser (AlfaVakuo, AS-Sr-5V-600) placed inside an under-vacuum ($\approx10^{-5}$ mbar) spectroscopy cell, similar in design to our previous work \cite{Cipris2024}. 
The dispensers are aligned along the z-axis, with the atomic beam emerging from the dispenser approximately 3 mm below the x-y plane.

A diode-pumped Yb:KGW femtosecond (fs) laser (Light Conversion, Carbide) is used to excite the atoms.
It propagates along the y-axis, perpendicular to the atomic beam, and is focused at the center of the cell.

To estimate the atomic velocity distribution at the point where the femtosecond laser intersects the atomic beam, we directed a continuous-wave (CW) laser at 461 nm (Moglabs, CEL) along the same path as the fs laser. 
We measured its transmission as a function of frequency by scanning over the $^1S_0 - ^1P_1$ transition, i.e., we measured the absorption spectrum. 
The resulting atomic velocity distribution closely resembles a Maxwell-Boltzmann distribution, with an estimated full width at half maximum (FWHM) of approximately $\approx 1$ GHz at a dispenser current of 9.5 A. 
Moreover, from transmission measurements using a cw laser and applying the Beer-Lambert law, we calculated the optical thickness of strontium atoms emerging from the dispensers at different currents.

The femtosecond (fs) laser operates at a nominal wavelength of $1030$ nm, with a pulse duration of $\approx 280$ fs and a repetition rate of $1$ MHz. 
To achieve the required excitation wavelength of 461 nm, the fs pulses are directed through a hybrid optical parametric amplifier (OPA, ORPHEUS-F-NS), followed by a pulse compressor and a second-harmonic generator (SHG). 
At the SHG output, a fs laser spectrum centered at 461 nm with a FWHM of 17 nm was measured. 
Due to this broad laser bandwidth, most of the photons are off-resonant with the $^1S_0$ - $^1P_1$ atomic transition at 461 nm.
The spectrum was further filtered to a FWHM of 1.6 nm using an optical grating (Thorlabs, GH13-24V) combined with a vertical slit (Thorlabs, VA100CP/M). 

The fs beam has elliptical profile at the focus with $1/e^2$ diameters of $0.31$ mm and $0.25$ mm along the x- and z- axes, respectively, as measured using a beam profiler (Ophir Optronics Sol., BM-USB-SP928-OSI).
For the average power of $36~\upmu$W used in all measurements (unless otherwise noted), and assuming the pulse has a Gaussian envelope, along with the mean beam diameter and a pulse duration of 280 fs, we find a pulse area of $0.08\pi$ \cite{Felinto2003}. 
During lifetime measurements, the power of the fs laser beam, as well as its focal position were continuously monitored using a beam profiler. 
The data showed that, across all lifetime measurements, the fs laser power fluctuated within $4 \%$ of the total power, while the focal point position drifted up to $4~\upmu$m along the x-axis and $22~\upmu$m along the z-axis.
While the measured changes are small, they do contribute to the background variations relevant for the lifetime measurements conducted over several days, as will be discussed in detail in Section \ref{sec: Analysis}.

The photons emitted through spontaneous decay were coupled into a 400~$\upmu$m diameter, multimode optical fiber (Thorlabs, M74L05) using a telescope with a 5:1 focal ratio, resulting in a 2 mm diameter imaging area.
The size of the imaging area was made as large as possible so as to minimize transit-time broadening.
Even atoms with velocities of 1000 m/s, a factor of $\approx2$ more than the most likely velocity in the atomic beam, would take 2~$\upmu$s to cross the imaging area, which is two orders of magnitude longer than the time span of our decay curve. This means that the effect of transit-time broadening on our measurements is negligible.
The optical fiber is 5 meter long and guides the photons to a hybrid single-photon detector (Becker $\&$ Hickl, HPM-100-07). 
The detector is characterized by extremely low afterplusing probability and quantum efficiency of $\approx15\%$ at 461 nm. 
During lifetime measurements, the temperature of the detector was monitored using a temperature data logger (Pico Tech., TC-08). 
For all lifetime measurements performed, the detector temperature variations were less than $0.7~^{\circ}$C, ensuring that the dark counts remained constant throughout the measurements.
For efficient single photon arrival timing, a SPC-130-EMN TCSPC module with a timing precision of $<3.5$ ps and a dead time of 100 ns was used. 
To suppress stray light, such as room light or fs laser light of other wavelengths used in the OPA and SHG, a bandpass filter centered at 460 nm (Thorlabs, FBH460-10) with a 10 nm FWHM was used in front of the detector.

In our measurements we use the reverse start-stop mode of TCSPC. In this mode, the time-to-amplitude converter (TAC) starts when a photon is detected and stops upon receiving a SYNC pulse from the laser. 
To ensure that the pulse exciting the atoms and producing the photon that starts the TAC is the same pulse that provides the SYNC to stop the TAC, we adjusted the length of the SYNC cable to introduce a 1~$\upmu$s delay, equivalent to the time between two consecutive fs pulses. 
This allows us to measure the arrival time of the detected photon relative to the laser pulse that caused the excitation which eliminates the effects of the fs pulse-period jitter.

All measurements were conducted within a time window of 66.025 ns, determined by adjusting the lengths of the Constant Fraction Discriminator (CFD) cable and the SYNC signal cable, and the TAC settings.
This configuration, with the addition of the dead time of the detector of 100 ns, leaves more than 800 ns of detector readiness for the arrival of the next photon. 
As a result, the influence of the detector's dead time was entirely eliminated.
The analog-to-digital converter (ADC) was set to its highest available resolution of 4096 bins, which provided a time-bin duration of 16.12 ps per bin.

Three pairs of bias coils were placed in three orthogonal directions (x, y, z) around the cell to compensate any stray magnetic field. 
The coils were significantly larger than the laser's interaction region with the strontium atoms, ensuring a homogeneous magnetic field within that area.
The magnetic field was measured at multiple locations around the cell using a magnetometer (Sensys GmbH, FGM3D/1000).
The field in the center of the cell, where the fs laser interacts with the atoms, was estimated through interpolation.
The magnetic field was reduced to zero with an error margin of 0.01 G, constrained by the sensitivity of the magnetometer.
In all measurements, the magnetic field was set to zero unless the bias coils were intentionally used to generate a non-zero DC magnetic field.
This non-zero field was applied to investigate the dependence of the atomic lifetime on the magnetic field, as detailed in Section \ref{sec:magnetic field}.

For a given set of experimental parameters (Sr dispenser current, external magnetic field, fs laser power, and focal position), measurements were conducted in the following order: first, the background was measured; then the signal, i.e., spontaneously emitted photons; and finally, the background was measured again.

We measure the background signal with the fs laser tuned to $\approx 457$ nm, i.e., sufficiently detuned from the $^1S_0$-$^1P_1$ transition to avoid excitation of the atoms.
Therefore, it includes both dark counts, and photons scattered and reflected from the interior and viewports of the spectroscopy cell, including dispensers.
Due to multiple reflections, these photons can reach the detector well after the fs excitation pulse has ended; in our case, their presence is detectable in the background signal for up to approximately 7.5 ns.
In each measurement sequence, we ensure that the total background signal measurement time (taken before and after signal) matches the duration of the on-resonance signal measurement.
Measuring the background signal before and after the on-resonance signal enabled us to estimate the error in the lifetime measurement caused by background fluctuations, which can arise from slight variations in the fs laser intensity and focal position.

%%%%%%%%%%%%%%%%%%%%%%%%%%%%%%%%%%%%%%%%%%%%%%%%%%%%%%%%%%%%%%%%%%%%%%%%%%%%%%%%%%%%%%%%%%%%%%%%%%%%%%%%%%%%%%%%%%%%%%%%%%%%%%%%
%%%%%%%%%%%%%%%%%%%%%%%%%%%%%%%%%%%%%%%%%%%%%%%%%%%%%%%%%%%%%%%%%%%%%%%%%%%%%%%%%%%%%%%%%%%%%%%%%%%%%%%%%%%%%%%%%%%%%%%%%%%%%%%%

\section{\label{sec: Analysis} Data analysis\protect\\}

\textit{Pulse pile-up correction}. Each measured dataset (both background and signal) is first corrected for pulse pile-up, which occurs when more than one spontaneously emitted photon reaches the detector within a single excitation cycle.
In such cases, the TCSPC registers the first photon but fails to register subsequent ones, which results in undercounting in later time bins.
The pile-up correction was made by multiplying the number of detected photon counts, $N_i$, in time bin $i$ by a factor defined as \cite{Gomez2005}:
\begin{equation}
F_{i} = \frac{1}{1-\frac{1}{N_E}\sum_{j<i-1}N_{j}},
\end{equation}
where $N_E$ is the number of excitation cycles and $N_j$ is the number of photon counts collected in the j-th time bin.
The largest correction factor, of approximately 1.028, is obtained for the signal measured at the highest strontium optical depth, where the photon counts is $\approx2.7 \%$ of the repetition rate.

\begin{figure}[h]
\includegraphics[width = \columnwidth]{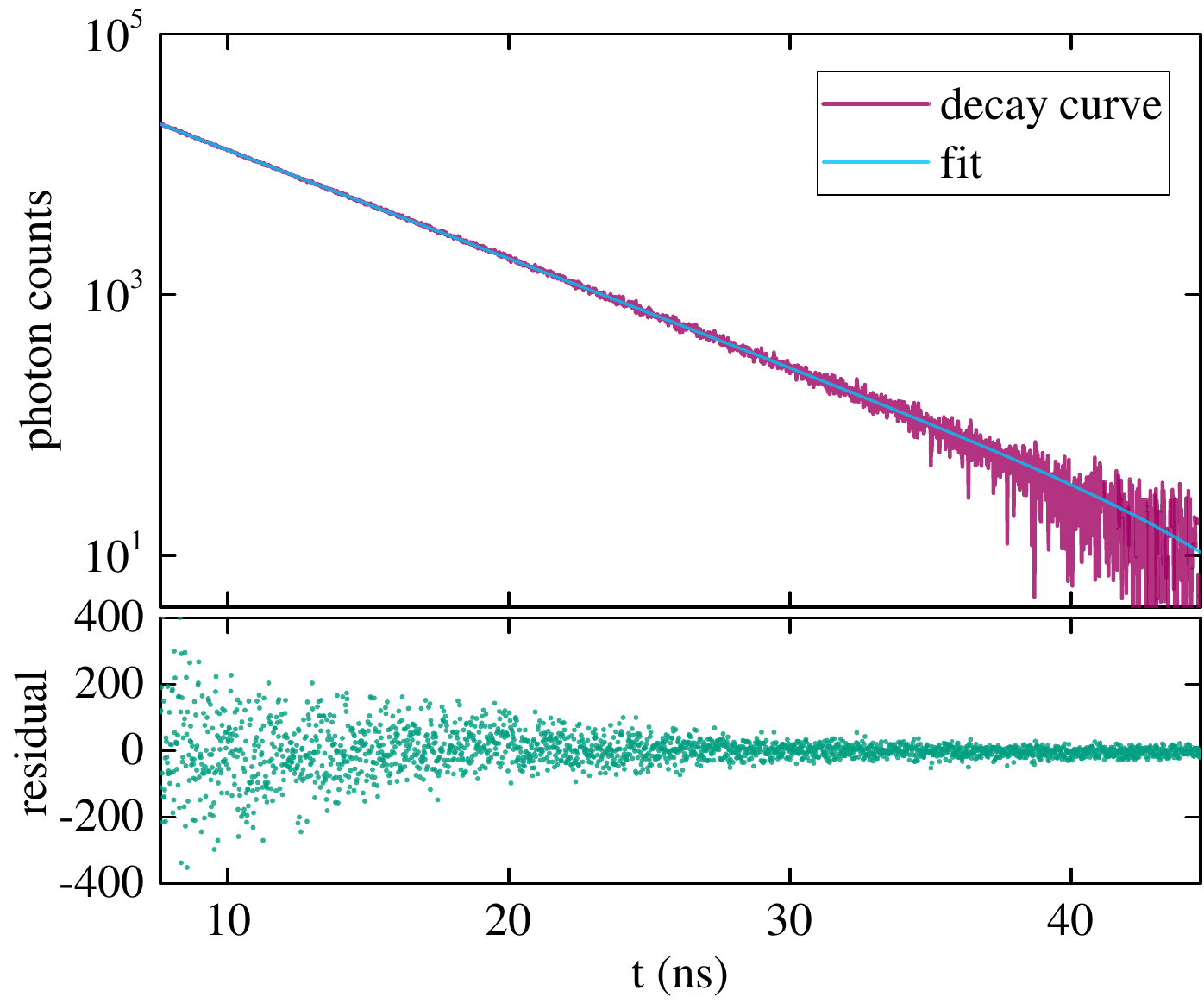}% Here is how to import EPS art
\caption{\label{fig:exp decay} Upper plot: Fluorescence decay curve of the $^1P_1$ state (pink) with the background signal removed. The corresponding decay curve that is the result of our fitting procedure is shown in blue. Note that the fitting function also includes the background signal which is not shown here. See text for more details about the fitting procedure. 
Lower plot: Fit residuals. }
\end{figure}

\textit{Fitting}. For a given signal dataset, after correcting the collected photon counts for the pile-up effect, we fit an exponential function of the following form:
\begin{equation}
N(t) = A\exp(-t/\tau) + p(B_1(t)+B_2(t)),
\label{eq: fit}
\end{equation}
where $\tau$ is the exponential decay time constant corresponding to the lifetime of the $^1P_1$ state, $A$ is the amplitude of the exponential function, $B_1(t)$ and $B_2(t)$ are background signals measured before and after the decay signal, respectively.
The total background signal, $B_1(t)+B_2(t)$, accounts for the dark counts, and photons reflected from the interior of the spectroscopy cell, including dispensers. 
We introduce the fitting parameter $p$ that multiplies the background signals to account for any changes during the acquisition time, coming mainly from drifts in laser power and pointing stability.
However, these changes are typically small, with p deviating within only a few percent from 1 in all measurements.

An example of a decay signal is shown in Fig.~\ref{fig:exp decay}, upper panel, where the background signal, multiplied by the fitting parameter $p$, has been subtracted.
The exponential decay part of the fitted function, $ A\exp(-t/\tau)$, is shown in blue.
The residuals are shown in the lower plot and are consistent with the shot noise corresponding to the signal level.

Although the fitting curve, according to relation \ref{eq: fit}, also includes the background, we observed that the obtained lifetime $\tau$ significantly depends on the starting bin time of the fit. 
Determining the adequate start bin time of the fit, commonly referred to as the truncation time is not straightforward, and requires a systematic approach \cite{Toh2019,Simsarian1998}.
In our approach, we use the measured signal dataset to generate multiple datasets for the fit, each with a different truncation time. 
The difference in truncation times between two consecutive datasets is 0.5 ns. 
We then perform a global fit, simultaneously fitting all datasets, each with a different truncation time, to a single function given by Eq. \ref{eq: fit}, where the parameters $A$ and $p$ are shared across all datasets.
Thus we obtain a set of $\tau_i$ values from the fit, each corresponding to the same measured signal dataset, but using different truncation time for the fit.
We limit the fit to the range from the selected truncation time up to t=44.6 ns, the point at which the photon count in the signal corresponds to the dark count level.
The global fit code was run using Wolfram Mathematica \cite{multifit}.
We analyzed the obtained $\tau_i$ values for different truncation times, and found that when the truncation time is set between 7.6 ns and 10.6 ns, $\tau_i$ remains constant, indicating no dependence on the choice of the fit’s start time. 
The final $\tau$ for a given measured signal dataset is determined as the mean of $\tau_i$ values obtained from the global fit within this truncation interval.
To estimate the uncertainty in $\tau$ for a given measured dataset, we calculate the standard deviation of $\tau_i$ values across the selected truncation interval, which we refer to as the truncation uncertainty. 
We then add in quadrature the largest fit error within this interval, and the background error to determine the total uncertainty in $\tau$ for the measured dataset.
The background error is estimated using again Eq. \ref{eq: fit} as the fit function, however, now only $B_1$ or $B_2$ are used, rather than their sum $B_1+B_2$. 
We define the background error as the half-difference of $\tau_{B_1}$ and $\tau_{B_2}$ obtained in this way.
The background error reflects variations in $\tau$ caused by small changes in the background during signal measurement.

\section{\label{sec: Lifetime} Lifetime of the 5\lowercase{s}5\lowercase{p} $^1P_1$ state\protect\\}

\begin{figure}
\includegraphics[width = \columnwidth]{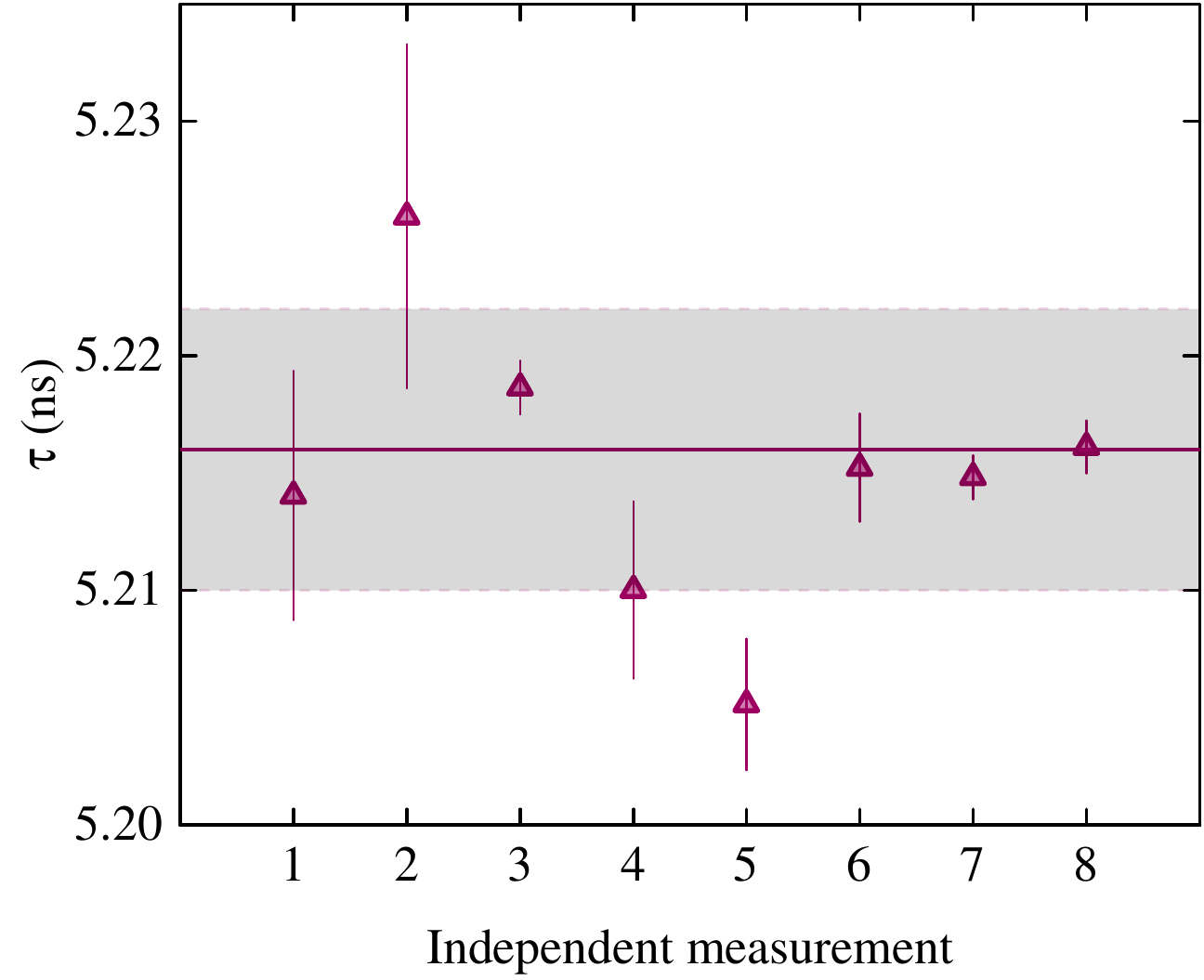}% Here is how to import EPS art
\caption{\label{fig: indep} Lifetime of the $^1P_1$ state determined from eight independent measurements. For data sets 1, 4 and 8, the signal was collected over 23, 38 and 20 hours, respectively, while the remaining data sets were collected over 15 hours. The error bars are calculated by adding the fit error, the truncation, and the background error in quadrature for each data set. The horizontal line represents the weighted mean of the 8 data sets, with the gray shaded area indicating the standard deviation. }
\end{figure}

In Fig. \ref{fig: indep}, we present the $^1P_1$ state lifetime obtained from eight independent measurements conducted under identical experimental conditions on different days over a two-month period.
For measurements denoted with numbers 1, 4 and 8, the signal was collected over 23, 38 and 20 hours, respectively, while the remaining datasets were collected over 15 hours. 
The optical depth of the atomic sample was 0.0011, estimated as described in Section \ref{sec:Exp setup}. 
The lifetime of the $^1P_1$ state and corresponding uncertainties were determined for each data set as explained in Section \ref{sec: Analysis}. 
The background error is the largest contributor to $\tau$ uncertainty for a particular data set. 
The average truncation error is $0.008\%$, while the maximum fitting error is $0.017\%$.

From these eight independent measurements and their uncertainties, the lifetime of the 5s5p $^1P_1$ state was determined by calculating the weighted mean, resulting in $\tau = 5.216$ ns, with a standard deviation, here referred to as the statistical error, of $0.006$ ns, see Fig. \ref{fig: indep}.

%%%%%%%%%%%%%%%%%%%%%%%%%%%%%%%%%%%%%%%%%%%%%%%%%%%%%%%%%%%%%%%%%%%%%%%

\section{\label{sec: Systematic effects} Systematic effects\protect\\}

Accurate determination of the lifetime of the $^1P_1$ state requires a  rigorous investigation of the systematic effects that could influence our result. 
We thoroughly examined all potential sources of uncertainty in $\tau$, including  signal nonuniformity, nonlinearity of the TAC, fs laser power, magnetic field, and radiation trapping. 

We summarize the contribution of these effects in the error budget in Table~\ref{tab:table1}.

After carefully considering each of these contributions, and including them into our result, we estimate the total uncertainty in the lifetime measurement to be $0.25\%$, which yields the final lifetime of the $^1P_1$ state of 

\begin{equation}
\tau(^1P_1) = (5.216 \pm 0.006_{stat} \pm 0.013_{sys})~\mathrm{ns}.     
\end{equation}

A detailed description of each contribution is provided below.

\begin{table}
\caption{\label{tab:table1}%
Error budget of the $^1P_1$ state lifetime measurement.
}
\begin{ruledtabular}
\begin{tabular}{lcdr}
\textrm{Error source}&
\multicolumn{1}{c}{\textrm{$\%$ Uncertainty}}\\
\colrule
TAC nonlinearity & 0.2  \\
Magnetic field &  0.004\\
Radiation trapping &  0.096\\
\colrule
Total sytematic error &   0.25\\
\colrule
Statistical error &  0.115\\
\end{tabular}
\end{ruledtabular}
\end{table}

\subsection{\label{sec: Time calibration} Signal nonuniformity \protect\\}
To test the signal nonuniformity of the ADC bins we collect scattered light from a CW laser at 461 nm. 
The laser power was adjusted so that the photon count rate matched the rate used in the lifetime measurements. 
Over an acquisition time of approximately 20 minutes, around 5000 photons per bin were collected. Since we are now using a CW source, i.e., a non-time-correlated source, we expect a uniform signal in all time bins. 
However, we find a small increase in counts with time. With a maximal time window of 1~$\upmu$s we find that the bins at the end of the window have on average 20 counts, or 0.4\%, more than the ones at the beginning, with an overall linear slope across the time window.  
We attribute this nonuniformity to time-correlated electronic noise in the signal acquisition system.
For the time window used for the decay rate data, 66.025~ns, this corresponds to 1.3 counts, two orders of magnitude lower than the typical dark noise level in our data. 
Therefore, we conclude that the signal nonuniformity does not contribute to the total error of our measurement.

\subsection{\label{sec: TAC nonlinearity} TAC nonlinearity\protect\\}

To determine time calibration errors in the TAC, two pulses - a start and a stop - were sent to the SPC-130-EMN TCSPC module using a digital delay generator (SRS, DG645). 
Additionally, the generated pulses were calibrated by a time interval counter (Keysight, 53220A). 
Varying the delay between the pulses while observing the TAC's output and applying a linear fit to the data, we observe a slope different from the expected one-to-one. 
The measured slope varies slightly from day to day, with the maximal observed slope being $1.00240(4)$. 
This indicates a small but measurable deviation from ideal linear behavior of the TAC. 
Effectively, this means that the slope of the decay curve would be smaller than measured, implying that the actual lifetime is shorter. 
To estimate the contribution to the uncertainty, we multiply time bins of a data set with the highest measured slope and fit using the mentioned fitting procedure.
We take the difference between this lifetime and the one obtained with no multiplication as the contribution to the uncertainty, equaling 0.2\% of the measured lifetime value.

\subsection{\label{sec:power} Laser power \protect\\}

Minor fluctuations in laser power during fluorescence decay measurements cause slight changes in the background during signal acquisition, affecting the accuracy of the fit used to determine $\tau$. This effect, referred to as the background error, is estimated for each individual dataset, as detailed in Section \ref{sec: Analysis}, and is included in the statistical error.

Another potential effect of laser power on lifetime could stem from spectral line broadening and saturation. 
Although our measurements were performed at low laser powers, resulting in low photon count rates, we evaluated this contribution as well.
We measured the signals and corresponding backgrounds for four different fs laser power, keeping all other experimental parameters constant.
The fs laser powers were 7$~\upmu$W, 36$~\upmu$W, 66$~\upmu$W and 68$~\upmu$W.
Using the measured data and following the procedure described above, we determined $\tau$ for each dataset measured at different fs laser powers. 
The uncertainty for each $\tau$ value was calculated by combining truncation, fit and background errors in quadrature. 
All obtained $\tau$ values fall within the calculated statistical uncertainty range, $\tau=5.216\pm0.006$~ns, indicating no dependence of $\tau$ on laser power within the power range used.

\subsection{\label{sec:magnetic field} Magnetic field \protect\\}

\begin{figure}[b]
\includegraphics[width = \columnwidth]{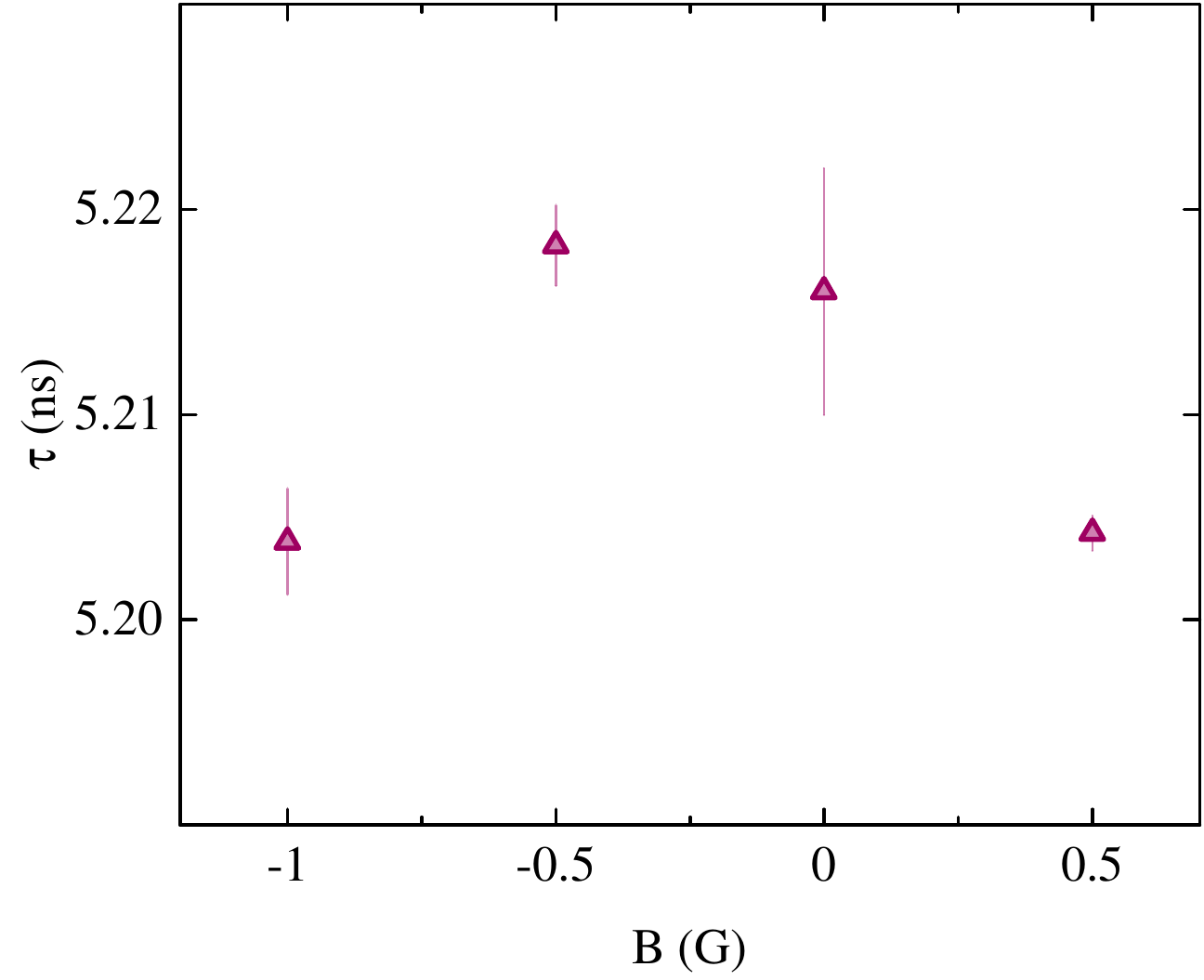}% Here is how to import EPS art
\caption{\label{fig: magfield} The measured lifetime of the $^1P_1$ state for various external magnetic field strengths.}
\end{figure}

In the presence of an external static magnetic field, the $^1P_1$ state splits into three Zeeman components, $m_J=0,\pm 1$.
The splitting between these components is given by $E_{ZE} = g_{J}\mu_{B}M_{J}B$, where $g_J$ is the Landé g-factor, $\mu_B$ is the Bohr magneton, $m_J$ is the magnetic quantum number, and B is the magnetic field strength, and has a value of 1.4 MHz/G. 

Given the spectral width of the fs laser, coherent excitation of the three Zeeman components of the $^1P_1$ excited state is possible, potentially leading to interference among the different $^1P_1 (m_J = \pm 1, 0) - ^1S_0 (m_J=0)$ decay paths.
This effect is known as quantum beats and causes the exponential decay function to be modulated by a cosine function \cite{Toh2019, BRANDENBERGER1981}, making Eq. \ref{eq: fit} no longer a suitable fit for the measured data.
The frequency of the modulation cosine function is equal to $E_{ZE}/h$, where $h$ is Planck's constant.

In our experimental setup, the magnetic field at the center of the cell, within the interaction zone of atoms and the fs laser, is zeroed to an accuracy of 0.01 Gauss, see Section \ref{sec:Exp setup}. 
At this field strength, the period of the cosine function which modulates the exponential function due to quantum beats is 36~$\upmu$s, which is four orders of magnitude longer than the measured time window relevant for the determination of the $^1P_1$ state lifetime. 
Thus, we estimate that for measurements performed at 0 Gauss, the effect of the modulation of the exponential fitting function by cosine function is negligible, resulting in no added uncertainty in $\tau$ related to the quantum beat effects.

To confirm this conclusion, we introduced an external static magnetic field to the interaction zone of the fs laser and strontium atoms and measured the dependence of $\tau$ on the applied magnetic field, keeping all other experimental parameters constant.
The external magnetic field was introduced along the y-axis, i.e., parallel to the laser propagation vector, using a pair of coils described in Sec. \ref{sec: exp}. 
Using the measured data and the global fitting procedure described above, we determined $\tau$ for each dataset measured at different magnetic field strength. 
The uncertainty for each $\tau$ value was calculated by combining    truncation, background, and fit errors in quadrature. 
Figure \ref{fig: magfield} presents the results, showing a certain dependence of $\tau$ on the strength of the external magnetic field; however, this dependence does not follow a straightforward functional form.

To conservatively estimate the uncertainty in $\tau$ near B=0, we used the largest observed deviation in $\tau$ across the measured magnetic field range. 
The largest deviation, measured at B=0.5 Gauss, is $0.19\%$ relative to $\tau$ at B=0. 
From this, we estimated that the deviation at B=0.01 Gaus, corresponding to the accuracy of the magnetic field measured at B=0, is approximately $0.004\%$. 
Although this is negligible compared to other sources of $\tau$ uncertainty, we have included it in the error budget table.

\subsection{\label{sec: radiation trapping} Radiation trapping \protect\\}

\begin{figure}[b]
\includegraphics[width = \columnwidth]{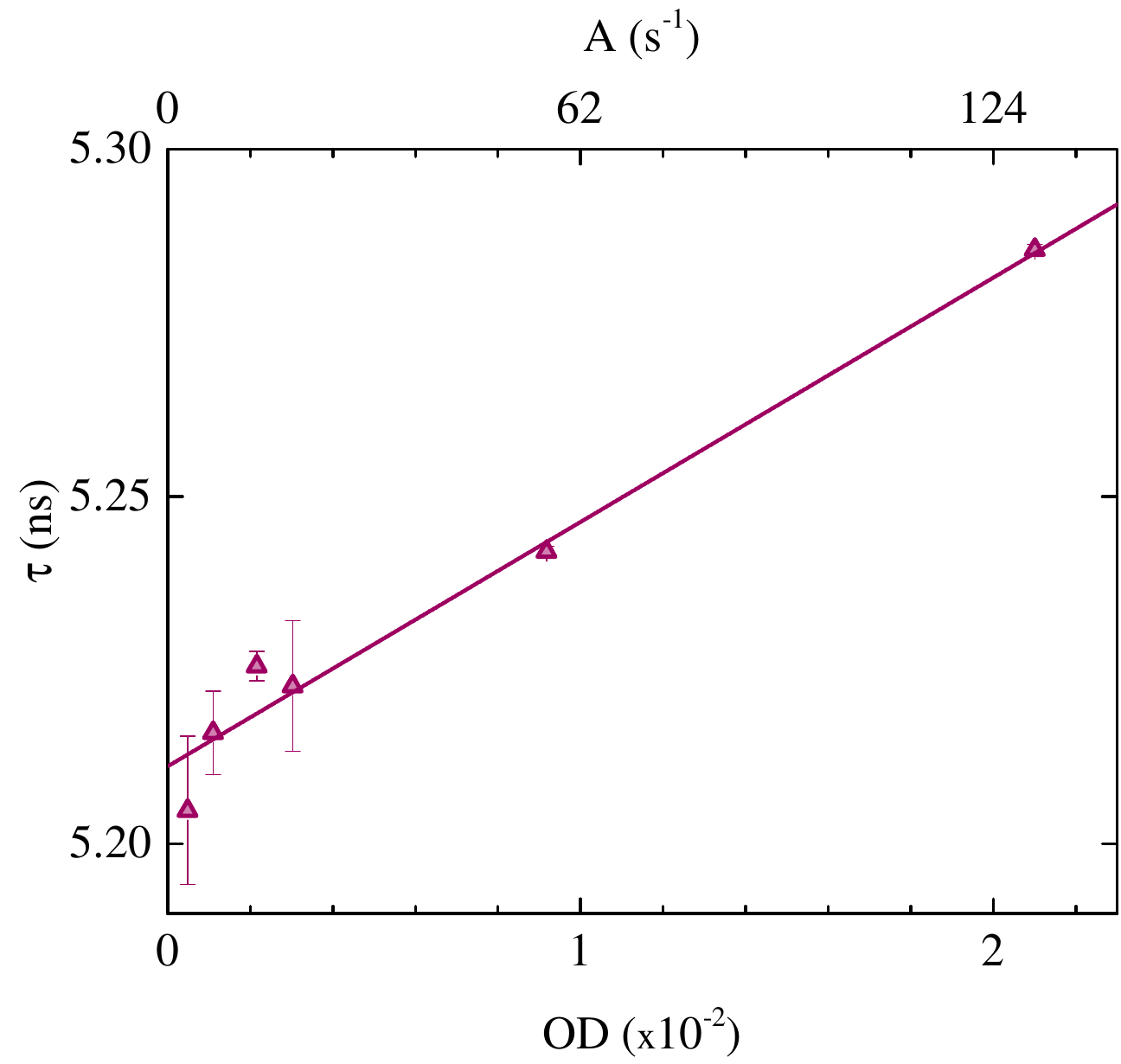}% Here is how to import EPS art
\caption{\label{fig: radtrapping} The measured lifetime of the $^1P_1$ state across a range of strontium optical depths (symbols). The solid line is a linear fit to data and yields an extrapolated lifetime of 5.211 ns for zero Sr density, with a fitting uncertainty of 0.002 ns.
The corresponding parameter $A$ is given on the upper x-axis}
\end{figure}

In dense vapors, the observed lifetime increases due to radiation trapping \cite{Milne1926, Patterson2015, Pucher2020}. 
To evaluate the influence of this effect, we measured the lifetime of the $^1P_1$ state at various optical depths (OD) of strontium vapor in the cell. 
We adjusted the OD by changing the current through the dispenser.
OD measurements were performed using the aforementioned cw laser at 461 nm.
For a given dispenser current, we recorded the transmission of the cw laser as a function of its frequency by scanning across the $^1S_0 - ^1P_1$ transition, effectively measuring the absorption spectrum. 
From this data, we calculated the OD for each current setting using the Lambert-Beer law.
For dispenser currents below 8.5A, direct OD measurements were challenging due to low absorption resulting from the low strontium atom concentration.
In these cases, we estimated the OD from the fit parameter $A$, which was confirmed to correlate with the concentration of strontium atoms.
To quantify this, we plotted $A$ as a function of OD for the measurable OD range, obtaining a linear relationship. 
These fit parameters were then used to calculate OD from the photon count rate for cases with lower dispenser currents.

Figure ~\ref{fig: radtrapping} shows measurements of $\tau$ as a function of OD. 
For completeness, the corresponding parameter $A$ is given on the upper x-axis.
A linear dependence of $\tau$ vs OD was found, consistent with previous literature \cite{Milne1926, Patterson2015, Pucher2020}.
The linear fit to data yields an extrapolated lifetime of 5.211 ns for zero Sr density, with a fitting uncertainty of 0.002 ns.
This results in a correction of $0.096\%$ for the lifetime of 5.216 ns, measured at OD = 0.0011 from eight independent measurements.

\section{\label{sec: Conclusion} Conclusion \protect\\}

To conclude, we have measured the lifetime of the $^1P_1$ state in Sr using the TSCPC technique. 
This work is the first such measurement of the lifetime of this state in strontium.
Additionally, we carefully analyzed systematic effects that effect our measurements. 
The largest contribution to the uncertainty, the TAC nonlinearity, could be drastically reduced by using another type of time-to-digital converter, for instance an FPGA \cite{Song2006}.
Nevertheless, our final result, $\tau(^1P_1) = (5.216 \pm 0.006_{stat} \pm 0.012_{sys})~\mathrm{ns}$, has a comparable error to previous measurements \cite{Blatt2020, Katori2006}.
As mentioned, these two previous measurements show a 7$\sigma$ discrepancy.
Our result agrees within $\sigma$ with the result obtained by Heinz et al. \cite{Blatt2020} from a tune-out frequency measurement, providing new and valuable information to the issue of the discrepancy with the lifetime determined from photoassociative spectroscopy \cite{Katori2006}.
We expect that this new spectroscopic data will improve the accuracy of calculations of the internal structure of the strontium atom, ultimately leading to more accurate strontium atomic clocks.

\section{\label{sec: Data} Data availability\protect\\}

The raw data and file description are available at \cite{DVN/92MET3_2025}.

\section{\label{sec: Funding} Funding\protect\\}
This work was supported by the following projects: New Imaging and control Solutions for Quantum processors and metrology - NImSoQ, funded through the QuantERA 2021 call (cofunded by the Croatian Science Foundation, HrZZ); Croatian Quantum Communication Infrastructure – CroQCI, funded through Digital Europe Call (Project Number: 101091513 and NPOO.C3.2.R2-12.01.0001); Centre for Advanced Laser Techniques (CALT), co-funded by the European Union through the European Regional Development Fund under the Competitiveness and Cohesion Operational Programme (Grant No. KK.01.1.1.05.0001).

\section{\label{sec: Acknowledgements} Acknowledgements\protect\\}
We thank G. Zgrabli\'c and S. Vdovi\'c for their support in operating the fs laser.

\providecommand{\noopsort}[1]{}\providecommand{\singleletter}[1]{#1}%

\bibliography{Puljic_etal_2025}

\end{document}